\documentstyle[floats,prl,aps]{revtex}
\input epsf.sty
\let\simgt=\gtrsim
\let\simlt=\lesssim
\begin{document}
\title{
Observationally Determining the Properties of Dark Matter}

\author{Wayne Hu\footnote{Alfred P. Sloan Fellow; revised \today}, 
Daniel J.\ Eisenstein, and Max Tegmark\footnote{Hubble Fellow}}
\address{Institute for Advanced Study, Princeton, NJ 08540}
\author{Martin White}
\address{Departments of Astronomy and Physics, University of Illinois
at Urbana-Champaign, Urbana, IL 61801}
\maketitle
\begin{abstract}
Determining the properties of the dark components of the universe remains 
one of the outstanding challenges in cosmology. We explore how upcoming 
CMB anisotropy measurements, galaxy power spectrum data, and supernova 
(SN) distance measurements can observationally constrain their gravitational 
properties with minimal assumptions on the theoretical side.  SN observations 
currently suggest the existence of dark matter with an exotic equation of 
state $p/\rho \simlt -1/3$ that accelerates the expansion of the universe. 
When combined with CMB anisotropy measurements, SN or galaxy survey data  
can in principle determine the equation of state and density of this component 
separately, regardless of their value, as long as the universe is spatially flat.  
Combining these pairs creates a sharp consistency check.  If $p/\rho \simgt -1/2$, 
then the clustering behavior (sound speed) of the dark component can be determined 
so as to test the scalar-field ``quintessence'' hypothesis.  If the exotic matter 
turns out instead to be simply a cosmological constant ($p/\rho = -1$), the 
combination of CMB and galaxy survey data should provide a significant detection 
of the remaining dark matter, the neutrino background radiation (NBR).  The gross 
effect of its density or temperature on the expansion rate is ill-constrained 
as it is can be mimicked by a change in the matter density.  However, anisotropies 
of the NBR break this degeneracy and should be detectable by upcoming experiments.
\end{abstract}

\section{Introduction}
\label{sec:introduction}

The nature of the dark matter remains one of the greatest outstanding
puzzles in cosmology.  
The difficulty of the problem is compounded by the fact that 
the dark matter may be composed of multiple components.
Despite this, we may be on the verge of an observational solution.  
The cosmic microwave background (CMB) contains information about the
dark components present in the early universe, specifically 
the ratio of
non-relativistic or cold dark matter (CDM) to relativistic species such as the
neutrino background radiation (NBR) and the ratio of the baryonic 
dark matter to the CMB itself.  
Upcoming high precision measurements of the CMB,
notably by 
the MAP \cite{Map} and Planck \cite{Planck} satellites, should determine these ratios 
to the percent level \cite{Jun96}.
In contrast, observations of high-redshift objects such as Type Ia supernovae (SN) probe dark components
important in the local universe.
Indeed preliminary results suggest the presence of
an additional dark component 
that accelerates the expansion \cite{Per97,Gar97}.
The clustering properties of galaxies link the CMB and the
local universe through their 
dependence on both the initial perturbations visible in
the CMB and the time-integrated history of structure formation between
last scattering and the present.  The galaxy power spectrum  
will be precisely measured by 
ongoing redshift surveys such as the 2dF \cite{2dF}
and the Sloan Digital Sky Survey (SDSS) \cite{SDSS}.

The promise of observationally
determining the properties of the dark components lies
in combining these data sets.  
Aside from the obvious difference in redshift windows, the
various data sets individually suffer from 
the fact that their observables 
depend degenerately on several aspects of the cosmology.  
Combined, they break each other's
degeneracies.  With three data sets, consistency tests become possible.
These tests are valuable for investigating systematic errors in the data
sets and could potentially indicate that the current cosmological framework
is inadequate to describe the universe.

In this paper, we investigate how the combination of CMB 
anisotropy measurements, galaxy survey data, and SN luminosity
distance determinations 
can be used to determine the parameters of the dark components. 
For this purpose, we employ the generalized dark matter
(GDM) parameterization scheme  
introduced in \cite{Hu98}. 
This parameterization encapsulates the observable properties
of the dark components in a background equation of state, 
its density today, a sound speed, and an anisotropic stress 
or ``viscosity parameter''.  
We begin by examining how well the equation of state and density today
can
be determined from observations assuming a flat universe.   
Combining
CMB data with either galaxy surveys or SN observations 
will provide
tight constraints on the equation of state and the density of the exotic
component even if the sound speed or viscosity must be simultaneously 
determined.  
The combination of these pairs will thus
provide a sharp consistency test.    
Galaxy survey information assists in these measurements indirectly
by freeing CMB determinations from
parameter degeneracies.  Its power is revealed 
only upon a full joint analysis and cannot be assessed
by direct examination of individual parameter variations
(c.f. \cite{Hue98}).  The fundamental assumption is that the
galaxy and matter power spectra are proportional on large
scales where the fluctuations are still linear. 

As the cosmological constant has a well-defined equation of state
$p/\rho = -1$, these cosmological measurements will test for its
presence.  A cosmological constant is also special because its density
remains smooth throughout the gravitational instability process.  If
the exotic component proves not to be a cosmological constant, then its
clustering properties become important.  These properties are
encapsulated in the sound speed.  The simplest models for this
component involve a (slowly-rolling) scalar-field ``quintessence''
\cite{Rat88,Sug92,Cob97,Cal98} which has the interesting property of
having the sound speed in its rest frame equal to the speed of light
\cite{Kod84,Hu98}.  By measuring the sound speed, one tests the
scalar-field hypothesis.  As long as $p/\rho \simgt -1/2$ in the exotic
component, the combination of CMB experiments and galaxy surveys can
provide interesting constraints on the sound speed.  Finally, we show
that if the exotic component turns out to be simply a cosmological
constant, then the properties of the remaining dark matter, the NBR, 
can be determined from combining
CMB and galaxy survey data.  In particular, anisotropies in the NBR, as
modeled by the viscosity parameter of GDM, are measurable and provide
a way to break the matter-radiation density degeneracy in the CMB.

The outline of the paper is as follows.  We
review the phenomenology of the
parameterized dark matter model of \cite{Hu98} in \S
\ref{sec:gdm} and the Fisher matrix technique for parameter 
estimation in \S \ref{sec:parameters}.  
In \S \ref{sec:eos}, we determine how well the equation of state
of the dark component may be isolated from its density. 
We discuss measurements of the sound
speed and propose a test of the scalar-field quintessence hypothesis 
in \S \ref{sec:sound}.  
In \S \ref{sec:nbr}, we address the detectability of the
NBR and its anisotropies.
We summarize our conclusions in \S \ref{sec:conclusions}.

\section{Dark Matter Phenomenology}
\label{sec:gdm}

In this section, we review the observable properties of the dark
components following the phenomenological treatment of
\cite{Hu98}.  We shall see that the properties of the dark
sector affect CMB anisotropies, structure formation, and high-redshift
observations in complementary ways.

We define the dark sector to include all components of matter that
interact with ordinary matter (baryons and photons) only gravitationally.
Thus, the observable properties of the dark sector 
are specified completely once the full stress-energy tensor is known.  Here we
consider the dark sector to be composed of 
background radiation from 3 species of essentially massless ($m_\nu \simlt 0.1$eV) 
neutrinos, 
a cold dark matter (CDM) component,
and an unknown component of generalized dark matter (GDM) \cite{Hu98}.
The CDM is required to explain the dynamical measures of 
the dark matter associated with galaxies and clusters.  The GDM
is required to be smooth on
small scales to avoid these constraints \cite{Tur97,Chi97}.
We further assume these forms of dark matter do not interact at the
redshifts of interest.  

Since each non-interacting species is covariantly conserved, 
the ten degrees of freedom of the symmetric stress-energy tensor of the
GDM are reduced to six.  We take these as the six 
components of the symmetric 3$\times$3 stress
tensor.   Two stresses generate vorticity and two generate gravity waves;
we will not consider these further but note that they may play a significant
role in CMB anisotropy formation in so-called ``active'' models for 
structure formation \cite{defects}.  This leaves two stresses: the isotropic
component (pressure) and an anisotropic component (``viscosity'').   The properties
of these two stresses must be parameterized.  We begin by 
discussing their effect on
the background expansion and then examine
their role in the gravitational instability of perturbations.

\subsection{Background Effects}
\label{sec:background}

\begin{figure}
\centerline{ \epsfxsize=5.25truein \epsffile{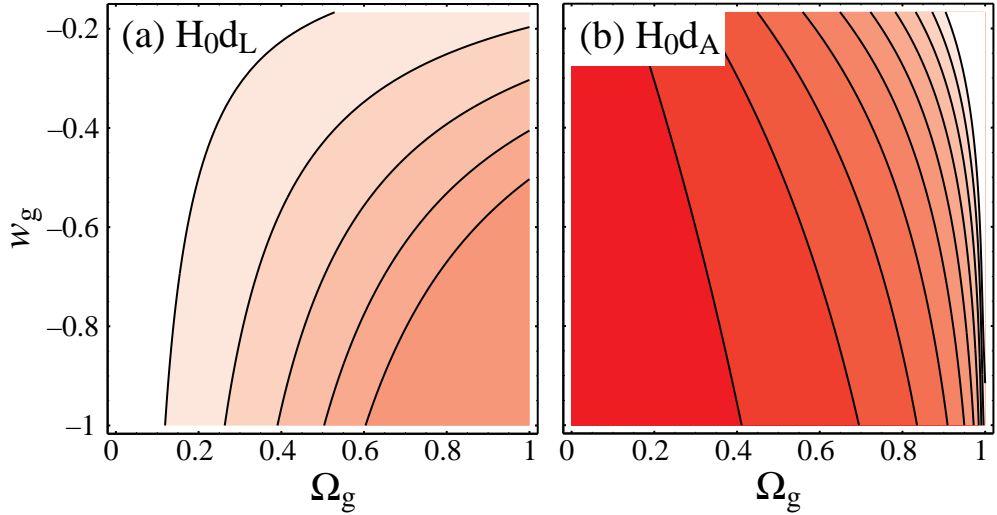}}
\caption{Distance measure degeneracies: contours of constant 
(a) luminosity distance to $z=0.5$ ($H_0 d_L$) and 
(b) angular diameter distance to
the last scattering surface ($H_0 d_A$).
$\Omega_m h^2$ has been held fixed in the former under the assumption that the CMB acoustic peak
morphology will measure it independently.   With this assumption, the two distance measures
provide complementary information.}
\label{fig:distance}
\end{figure}

Isotropy demands that to lowest order 
the GDM stress tensor has only an isotropic (pressure) 
component.
The GDM properties to lowest order therefore depend only
on 
the equation of state $w_g = p_g/\rho_g$.  For example,
energy conservation requires
that the evolution of the
GDM density follows
\begin{equation}
{d \ln \rho_g \over d \ln (1+z)^{-1}}  =  -3 (1+w_g)\,. 
\end{equation}
For simplicity, we consider models where $w_g$ is independent
of the redshift $z$ since
the properties of a slowly-varying $w_g$ 
can be modeled over the relevant redshifts with an appropriately
weighted average \cite{Cal98,Whi98}.   
Combined with the assumption of zero spatial curvature, 
the expansion rate or Hubble parameter becomes
\begin{equation}
{H \over H_0}(\Omega_g, w_g, \Omega_m/\Omega_r; z) =  [\Omega_m (1+z)^3 + \Omega_{g} (1+z)^{3(1+w_g)} 
+ \Omega_r (1+z)^4]^{1/2}\,,
\label{eqn:hubble}
\end{equation}
where $H_0 = 100\, h$ km s$^{-1}$ Mpc$^{-1}$.
Here 
$\Omega_m = 1- \Omega_g - \Omega_r$ and 
accounts for both the CDM and baryonic (``matter'') 
components.  Likewise $\Omega_r$ accounts for
the photon and neutrino (``radiation'') components.  
Furthermore, quantities that depend on the redshift behavior of
the expansion rate such as the deceleration
\begin{equation}
q = -{(1+z)^{-1} \over H } {d H \over d (1+z)^{-1}} + 1
\end{equation}
depend on the same parameters.  In particular, a component with
$w_g < -1/3$ is able to drive $q$ negative and cause an acceleration.

\begin{figure}
\centerline{ \epsfxsize=3.5truein \epsffile{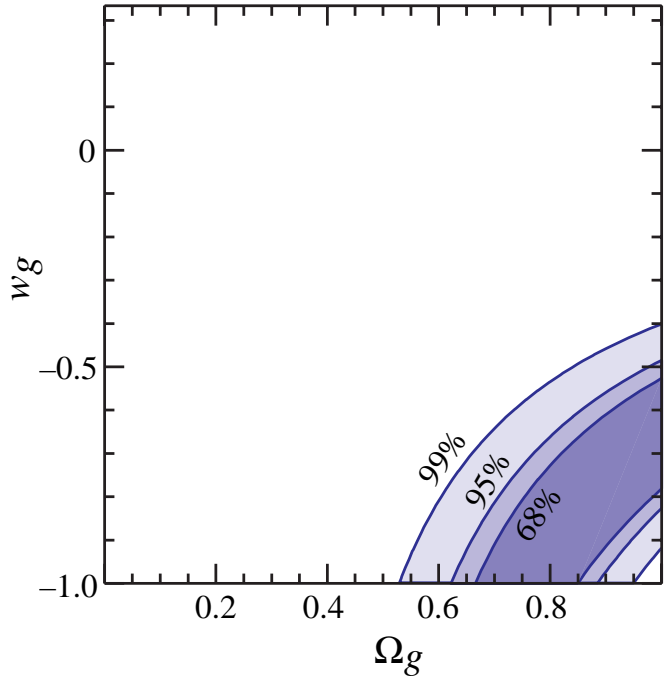}}
\caption{Current SN data.  
The 65\%, 95\% and 99\% CL intervals
in the $w_g-\Omega_g$ plane for the current data assuming only
statistical errors.  Constraints include 6 high redshift SN from
the Supernova Cosmology Project
\protect\cite{Per97} and 10 from 
the High-z Supernova
Search \protect\cite{Gar97}.  We use 26 low-$z$ calibrating
SN with $B-V<0.2$ obtained
by the Cal\'an/Tololo group \protect\cite{CalTol}. 
The analysis
follows \protect\cite{Whi98}, but note that systematic errors may dominate
in the current data sets.  
}
\label{fig:sncurrent}
\end{figure}

Any cosmological observable that is simply a function of
the expansion rate of the universe will only be sensitive to the background
properties of the GDM through $(w_g,\Omega_g)$.
For example, SN probe the luminosity distance (see Fig.~\ref{fig:distance}a)
\begin{equation}
H_0 d_L(w_g,\Omega_g;z) = (1+z) \int_0^z dz\, H_0/H\,;
\label{eqn:dl}
\end{equation}
this function is independent of the matter-radiation ratio because the observations
are at sufficiently low redshift.  
Current SN luminosity distance data suggest the presence of
an accelerating component with $w_g < -1/3$ (see Fig.~\ref{fig:sncurrent})
but may be dominated by unknown sources of systematic errors.

Likewise the acoustic peaks in the CMB probe the angular diameter distance 
to the redshift of
last scattering $z_{\rm ls}$  (see Fig.~\ref{fig:distance}b)
\begin{equation}
H_0 d_A(w_g,\Omega_g,\Omega_m/\Omega_r;z_{\rm ls}) = \int_0^{z_{\rm ls}} dz\, H_0/H\,,
\label{eqn:da}
\end{equation}
through its ratio with the sound horizon at last scattering 
$H_0 s_{\rm CMB}(\Omega_\gamma/\Omega_b, \Omega_m/\Omega_r)$.  Current CMB detections
alone do not place significant on $w_g$ and $\Omega_g$ \cite{Whi98}.

The energy density of the CMB is fixed through the measurement
of its temperature 
$T_{\rm FIRAS} =2.728\pm 0.004$K (95\% CL) \cite{Fix96} such that
\begin{equation}
\Omega_\gamma h^2 = 2.4815 \times 10^{-5} 
\left({T_{\rm CMB}\over T_{\rm FIRAS}}\right)^4  \,.
\end{equation}
With this constraint, the dependence of the sound horizon on 
the photon-to-baryon ratio reduces to a dependence on
$\Omega_b h^2$.
Likewise, the total radiation energy density is
given by
\begin{equation}
\Omega_r h^2 = \Omega_\gamma h^2
	\left[1 + 0.681 {N_\nu  \over 3} 
	\left(1.401 T_\nu \over T_{\rm CMB} \right)^4\right] \,.
\label{eqn:omeganh2}
\end{equation}
Thus 
under the
usual assumptions for the number of neutrino species ($N_\nu=3$)  
and their
thermal history ($T_\nu = T_{\rm CMB}/1.401$), 
the dependence of 
$d_A$ and $s_{\rm CMB}$ on $\Omega_m /\Omega_r $ becomes
a dependence on $\Omega_m h^2$.  We relax these 
assumptions in \S \ref{sec:nbr} to test the properties of the NBR.

\subsection{Structure Formation}
\label{sec:structure}

To probe the remaining properties of the GDM, one must consider
its effects on the gravitational instability of perturbations.  
Unless $w_g=-1$ (the cosmological constant case 
$\Omega_g=\Omega_\Lambda$),
the GDM participates in 
the gravitational instability process. 
The cosmological constant case is special since the relativistic
momentum density, which is proportional to $1+w_g$, goes to zero.

Since the perturbations need only
be statistically isotropic, the GDM in general requires two parameters to describe
fluctuations in its stress tensor.  These can be chosen to be the
sound speed in the rest frame of the GDM $c_{\rm eff}$, where 
$c^2_{\rm eff} \equiv
\delta p_g/\delta\rho_g$ (in units of where $c=1$), 
which relates the pressure fluctuation to the density
perturbation, and
a ``viscosity'' parameter $c_{\rm vis}$, which relates velocity 
and metric shear to the anisotropic stress. See \cite{Hu98} for their
precise covariant definition.
Positive values of $c^2_{\rm eff}$ imply that density fluctuations are
stabilized by pressure support at the effective sound horizon
$s_{\rm eff}\equiv \int c_{\rm eff} (1+z)dt$.  Likewise, positive values of
$c^2_{\rm vis}$ imply that resistance to shearing stabilizes the
fluctuation at $s_{\rm vis}\equiv \int c_{\rm vis} (1+z)dt$.  
These definitions assume that $c_{\rm eff}$ and $c_{\rm vis}$,
respectively, are slowly varying.
We call the greater of $s_{\rm eff}$ and $s_{\rm vis}$
the GDM sound horizon $s_{\rm GDM}$.

\begin{figure}
\centerline{ \epsfxsize=5.75truein \epsffile{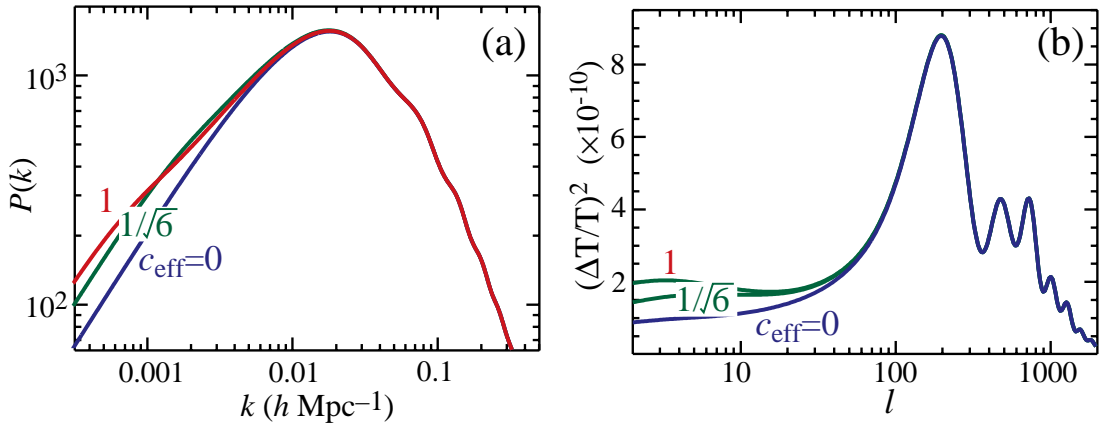}}
\caption{Sound speed effects for $w_g=-1/3$.  
A finite sound speed $c_{\rm eff}$ stabilizes perturbations
leading to features in the (a) galaxy power spectrum  and (b) CMB anisotropy spectrum.  
Note that in both cases the effects change most rapidly with $c_{\rm eff}$ between $0$ and
$\sqrt{1/6}$.  The power spectra have been normalized to small scales to bring out the
degeneracies and the importance of
large-scale information.   The model here and throughout has
$\Omega_m=0.35$, $h=0.65$, $\Omega_b h^2 = 0.02$, $\tau = 0.05$,
$n=1$, and $T/S=0$.
} 
\label{fig:psound}
\end{figure}

Modes smaller than $s_{\rm GDM}$ are stabilized by stress support.
If the GDM also dominates the expansion rate
then the growth of structure will slow below this scale.  
If $w_g < 0$, GDM domination occurs at approximately
\begin{equation}
1+z_g \equiv \left( { \Omega_g \over \Omega_m } \right)^{-1/3 w_g} \,.
\end{equation}
Thus we expect a feature in the matter power spectrum between
$H_0 s_{\rm GDM}(w_g,\Omega_g)$ at $z = z_g$ and $z=0$.   Since below the
sound horizon, the effect of GDM is to slow the growth of
structure independent of scale, the determination of $c_{\rm eff}$
from measurements of the galaxy power spectrum depends crucially
upon having data across this range of scales.  
In Fig.~\ref{fig:psound}a, we show the effect of varying $c_{\rm eff}$
on the power spectrum for $w_g=-1/3$.  The models have been normalized to small scales
to bring out the scale-independence of the small-scale suppression.
Given that the normalization is uncertain because of the unknown 
proportionality constant between the mass and galaxy power spectrum usually
defined as $b^2$ where $b$ is the ``bias''.
the only direct information from galaxy surveys on 
$c_{\rm eff}$ comes from large scales.  
Notice the most rapid change with $c_{\rm eff}$ 
occurs for $c_{\rm eff} < \sqrt{1/6}$.

The location of the GDM sound horizon also affects CMB anisotropies.
By halting the growth of structure, the GDM causes gravitational potential
wells to decay below the sound horizon 
$s_{\rm GDM}$ at $z_g$.   A changing gravitational potential
imprints fluctuations on the CMB via differential gravitational
redshifts whose sum is called the integrated Sachs-Wolfe (ISW) effect. 
However, if the sound horizon is much smaller than the 
particle horizon, the photons will traverse many wavelengths of the
fluctuation as the potential decays.  The cancellation of red and blue 
shifts destroys the effect.  Thus the largest effect arises
when $c_{\rm eff} \sim 1$ but varies strongly 
with $c_{\rm eff}$ only as it becomes less than $c_{\rm eff}
\sim \sqrt{1/6}$.  Unfortunately,
subtle differences in this large-angle 
temperature signal around $c_{\rm eff}=1$  will be difficult 
to pin down
given cosmic variance.

Because of the strong $c_{\rm eff}$-dependence of 
features in the CMB and galaxy survey power
spectra 
for $c_{\rm eff} \simlt \sqrt{1/6}$, tight lower limits can be placed
on $c_{\rm eff}$ from the data even though models around 
$c_{\rm eff}=1$ cannot be distinguished.   Furthermore, the
amplitude of the features decreases sharply as $w_g \rightarrow -1$
since even clustering above the sound horizon vanishes in this limit. 
Constraints on the sound speed will thus only be possible if the
equation of state of the GDM differs significantly from a 
cosmological constant.

\begin{figure}
\centerline{ \epsfxsize=4.25truein \epsffile{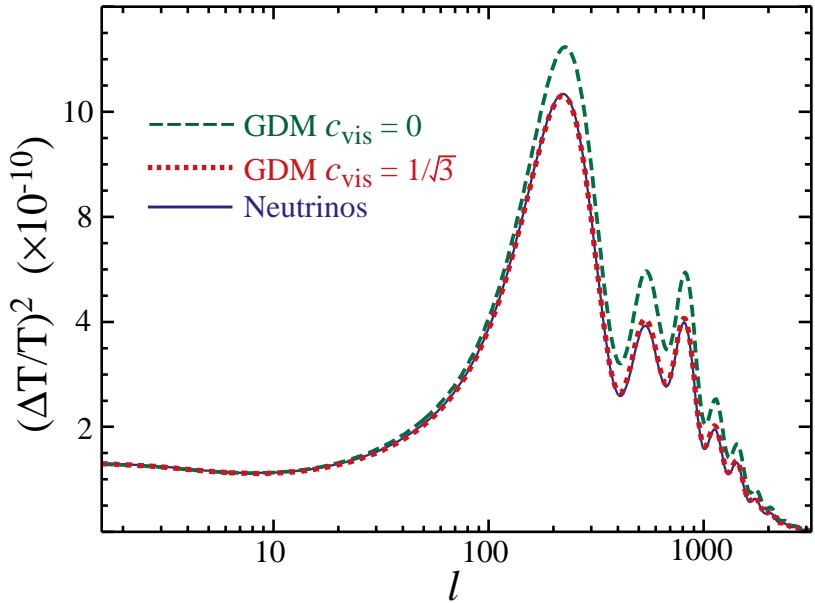}}
\caption{Viscosity effects: neutrinos may be accurately modeled as GDM with a viscosity
parameter $c_{\rm vis}=\sqrt{1/3}$.  Setting $c_{\rm vis}=0$ changes the CMB anisotropies significantly
and equate to removing the quadrupole anisotropy of the NBR. 
}
\label{fig:pcvis}
\end{figure}

Finally, the viscous term also causes the growth of 
structure to halt.  However,
it has an additional effect that makes it unique since anisotropic stresses 
enter directly in the Poisson equation that
 defines their relation to the gravitational potentials \cite{Hu98}.  
In Fig.~\ref{fig:pcvis}, we show
the effect of replacing the neutrinos with GDM of $c_{\rm vis}=\sqrt{1/3}$ and $0$ in the CMB.  
The former component models the neutrinos accurately and the latter shows that
the anisotropic stress of the dark matter produces potentially observable 
effects in the CMB.   We will
exploit this effect to propose a means of detecting the anisotropies in the neutrino
background radiation in \S \ref{sec:nbr}.

\section{Parameter Estimation}
\label{sec:parameters}

Projections for how well various data sets can measure cosmological parameters
depend crucially on the extent of the
parameter space considered as well as on the location in this space (or
``fiducial model'') around which we quote our errors.  Even though the latter uncertainty
will be eliminated once the best fit model is found from the actual
data, the former problem will remain.
The extended GDM
parameter space thus allows the data the freedom to choose the best
values to describe properties of the dark sector.  Even if the true
model turns out to contain only conventional dark matter, it allows us
to say with what confidence we can make this statement, i.e.~that the
dark components are in fact the neutrino background radiation, a
cosmological constant, and cold dark matter.  If these options are
ruled out, then we will have discovered a new form of matter.

We adopt a 10-dimensional parameterization of cosmology that includes
the present density of the GDM $\Omega_g$,
a time-independent equation of state $w_g$, the matter density $\Omega_m h^2$, 
the baryon density $\Omega_b h^2$,
the reionization optical depth $\tau$,
the tilt $n$, 
the tensor-scalar ratio $T/S$, 
the normalization $A$, 
and the linear bias $b$.
Both $c_{\rm eff}$ and $c_{\rm vis}$ affect the clustering 
scale and are largely degenerate. 
In \S \ref{sec:eos} and \S \ref{sec:sound}, we take $c_{\rm eff}$ as
a proxy for both, 
whereas in \S \ref{sec:nbr}, we take $c_{\rm vis}$ since we 
are interested in the anisotropy itself.
This parameter space does not include models with non-zero spatial
curvature or with massive neutrinos.
We take the fiducial model to have $\Omega_m=1-\Omega_g=0.35$, 
$h=0.65$, $\Omega_b h^2=0.02$, $\tau=0.05$, $n=1$, and $T/S=0$; we will use 
fiducial models with different values of $w_g$ and $c_{\rm eff}$
to explore how the results depend on these parameters.
For how the fiducial choices in the standard parameters affect the errors, see
\cite{Jun96,Eis98}.

To estimate errors on the cosmological parameters, we employ the Fisher
matrix formalism (see \cite{Teg97a} for a general review).
The Fisher matrix is essentially an expansion of the log-likelihood function
around its maximum in parameter space.  It codifies the optimal errors on each parameter 
for a given experiment assuming a quadratic approximation for this function.  
Note that the Fisher matrix provides accurate estimates 
of error contours only when they encompass small variations 
in parameter space where the expansion is valid.  This is
an important caveat to which we will return in \S \ref{sec:sound}.

Fisher matrix errors thus employ derivatives of the cosmological 
observables with
respect to parameters.  To calculate derivatives of the CMB and galaxy
power spectra, we employ the hierarchical Boltzmann code of \cite{Hu95,Whi96}.  
Special care must be taken in evaluating these derivatives 
since numerical noise in the
calculation can artificially break any parameter degeneracies that exist. 
General techniques such as the taking of two-sided derivatives and
the associated step sizes for standard parameters are
described in \cite{Eis98}.  For the GDM parameters, we estimate the
derivatives by finite differences with the step sizes
$\Delta \Omega_g = \pm 0.05(1-\Omega_g)$, $\Delta w_g = \pm 0.01$,
$\Delta c_{\rm eff}^2 = \pm 0.1 c_{\rm eff}^2$ and
$\Delta c_{\rm vis}^2 = \pm 0.1 c_{\rm vis}^2$.

The benefit of a hierarchy treatment is that unlike 
the integral treatment of CMBfast \cite{Sel96}, no interpolation is necessary and the
code may be made arbitrarily accurate 
by adjusting the sampling in Fourier space. 
Thus the  numerical noise problems identified by \cite{Eis98} can be addressed 
and in principle eliminated.
However, computational speed generally requires a compromise involving
smoothing the calculated CMB power spectrum \cite{Hu95}.
We have tested our results against CMBfast version~2.3.2 in 
the context of flat $\Lambda$-models where CMBfast is most accurate and 
obtained $\sim 5\%$ agreement in 
parameter estimation with the hierarchy code.  
The test also involved two independent pipelines for taking model calculations
through to parameter estimations.
These comparisons 
were done with 1500 Fourier modes out 
to $k=2.5k_{\rm damp}$, where $k_{\rm damp}$
is the CMB damping scale (see \cite{Hu97d}, Eq.~[17])
with Savitzky-Golay smoothing \cite{Pre92}
of the resulting CMB angular power spectrum.
We adopt the same techniques for parameter estimation in
the GDM context.\footnote{Note that in cases where strong model degeneracies
and high experimental sensitivity coexist, 2000 modes are often required.  
For this reason, we do not quote errors for the Planck experiment alone;
once combined with SN or galaxy survey data, its degeneracies are broken 
well enough that these small numerical errors are irrelevant.}

In addition to the extent of the parameter space and
the location of the fiducial model in parameter space, 
Fisher errors of course depend on the sensitivity of the given
experiment.  
For the CMB data sets, we take the specifications of the 
MAP and Planck experiments given in \cite{Eis98}.  
We quote results with and without polarization information.
Since the polarization signal may be dominated by foregrounds 
and systematic errors, the purely statistical Fisher matrix
errors may be underestimates.
For galaxy surveys, we take
the Bright Red Galaxy sample of SDSS; the specifications and their
translation into the Fisher matrix formalism is given in \cite{Teg97b}. We further 
take the linear power spectrum for parameter estimation. 
Because non-linear
effects and galaxy formation issues complicate the interpretation of
the observed power spectrum on small scales, we employ information only from
wavenumbers less than $k_{\rm max} = 0.2 h$Mpc$^{-1}$ and show how the
results change as we go to a more conservative $k_{\rm max}=0.1 h$Mpc$^{-1}$.  
This roughly brackets the regime where non-linear effects begin
to play a role as shown by simulations \cite{Mei98}.
For SN, we assume that on a timescale comparable to the MAP mission
a total of $200$ supernovae will be found with individual magnitude
errors of 0.3 and a redshift distribution with a mean of $z=0.65$ 
and a gaussian width of $\Delta z=0.3$ \cite{Teg98}.  

As each data set is independent to excellent approximation, 
the combined likelihood function
is the product of the CMB, SN and galaxy survey
likelihood functions.  Thus the combined Fisher matrix is simply
the sum of the individual Fisher matrices.

\section{Measuring the Equation of State}
\label{sec:eos}

Current SN luminosity distance measures suggest that
the GDM may have an exotic equation of state $w_g \simlt -1/3$ that
accelerates the expansion 
(see Fig.~\ref{fig:sncurrent} and \cite{Per97,Gar97}).  
If these preliminary indications
are borne out by future studies, one would like to pin down the
equation of state of the GDM and also construct consistency tests
to verify this explanation of the SN data.
Unfortunately, no one data set can isolate the equation of state on its own.   
The CMB has one
measure of $w_g$ from the angular diameter distance 
to the last scattering surface (see Eq.~[\ref{eqn:da}]),
but this is degenerate with $\Omega_g$ (see Fig.~\ref{fig:distance}),
even assuming zero curvature and that $\Omega_m h^2$ has been measured from the morphology of the
acoustic peaks.
Leverage on $w_g$ and $\Omega_g$ comes only through the
cosmic-variance-limited ISW effect at large angles and through the 
effects of gravitational lensing on very small scales.  The latter
occurs because changing $\Omega_g$ affects the present-day normalization
of the matter power spectrum and thereby changes the amount of lensing.
The effect is small, however, and much less powerful for breaking 
the degeneracy than the methods of the next paragraph; we therefore
neglect lensing in the CMB.
Likewise, galaxy
surveys alone give leverage only through the combination that defines
the GDM sound horizon.  For SN measurements that span only a short
range in redshift, there is an analogous 
degeneracy between $w_g$ and $\Omega_g$ in the luminosity 
distance (see Eq.~[\ref{eqn:dl}]). 

\begin{figure}
\centerline{ \epsfxsize=5.25truein \epsffile{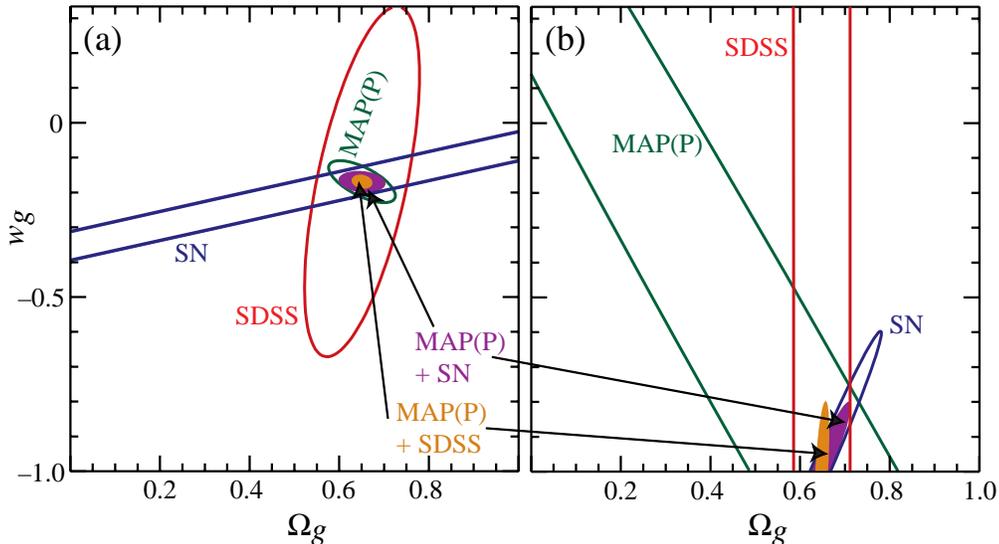}}
\caption{Breaking the equation-of-state density degeneracy.  The
degeneracy exposed in Fig.~\protect{\ref{fig:distance}} can be broken
by combining CMB information (here from MAP) with SN or galaxy survey
(here from SDSS) information.  Plotted here are the 68\% CL for the
various experiments and combinations.
Comparison of the two combinations leads to a sharp consistency test.  
Two fiducial models are shown: (a) $w_g=-1/6$ and (b) $w_g=-1$. 
While the CMB alone does well at $w_g=-1/6$, its degeneracy worsens
considerably as $w_g$ decreases toward $-1$.  Note, however, that
the complementary nature of the
data sets and the ability to make consistency checks occurs in both cases.}
\label{fig:wgomg}
\end{figure}

However, combining {\it two} of these measurements isolates $w_g$ and
the third can be used as a consistency check.  That the CMB angular
diameter distance and SN luminosity distance measures break each others
degeneracies is obvious from comparing panels (a) and (b) in Fig.\ 
\ref{fig:distance}.  We show in Fig. \ref{fig:wgomg} the error ellipses
(68\% CL\footnote{A 68\% confidence region for a two-dimensional
ellipse extends to $1.52\sigma$ along each axis.  We use this in
all cases, although it is an overestimate in cases when the ellipse
extends into unphysical or implausible regions of parameter space.}) 
in the $\Omega_g-w_g$ plane.  

Due to the large ISW effect in the
$w_g=-1/6$ model of Fig.~\ref{fig:wgomg}a, 
MAP alone will provide reasonable 
constraints on the two parameters.  
In this case, 
SN measurements will provide
a strong consistency check on CMB measurements. However,
 as one approaches $w_g=-1$, the ISW effect decreases and the
CMB requires the assistance of SN measurements to break the
degeneracy. 
We list the 1$\sigma$ errors as a function of $w_g$
in Tab. \ref{tab:wgomg}.

\begin{figure}
\centerline{ \epsfxsize=4.25truein \epsffile{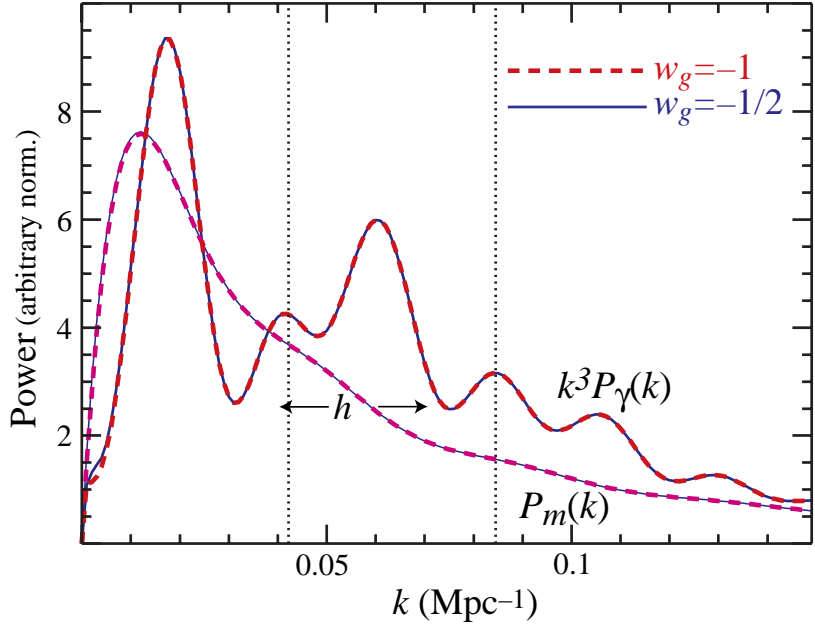}}
\caption{Measurement of $h$ and the equation of state degeneracy.
Acoustic features in both the CMB ($k^3 P_{\gamma}$) and matter/galaxy
power ($P_m$) spectra are frozen in at last scattering.  Once the CMB
acoustic peaks are calibrated in real space from the peak morphology
measurement of $\Omega_m h^2$ and $\Omega_b h^2$, sliding the galaxy
power spectrum in redshift space ($h$ Mpc$^{-1}$) until the features
``match'' determines $h$.  This test is unaffected by and therefore insensitive
to late-time dynamics from the GDM component.  
However, once $h$ is determined,
$\Omega_g=1-\Omega_m$ follows from the CMB measurement of $\Omega_m
h^2$.  The angular diameter distance measurement from the CMB then
determines $w_g$.}
\label{fig:trans}
\end{figure}

The combination of CMB and galaxy survey data provides a more
subtle example of complementarity.  Despite the fact that the MAP error
ellipse lies wholely within the SDSS error ellipse in
Fig.~\ref{fig:wgomg}a, the addition of SDSS provides substantially
smaller error bars.  As discussed in \cite{Eis98L}, the combination of
the CMB and galaxy power spectrum information yields a precise
measurement of the Hubble constant $h$ and $\Omega_m$ independently of
any low redshift GDM effects.  The reason is that the physical extent
of the sound horizon at recombination can be precisely calibrated from
measurement of $\Omega_b h^2$ and $\Omega_m h^2$ through the acoustic
peak morphology (see Fig.~\ref{fig:trans} and \cite{Hu96}).
Measurement of this scale in redshift space isolates the Hubble
constant; the aforementioned measurement of $\Omega_m h^2$ in the CMB
then returns $\Omega_m$.  Under the assumption of a flat universe,
$\Omega_g=1-\Omega_m$ is also well determined (see Fig.\ 
\ref{fig:wgomg}b) and the angular diameter distance $d_A(\Omega_m h^2,
\Omega_g, w_g)$ may be used to extract $w_g$.  
Hence, despite the fact that galaxy surveys cannot 
determine these parameters by themselves, they can break
the angular diameter distance degeneracy of the CMB and thereby
allow the CMB to measure $w_g$ and $\Omega_g$.
The subtle nature of the degeneracy breaking requires a full joint
analysis to uncover (c.f.\ \cite{Hue98} who reached more pessimistic
conclusions from a separate examination of each data set).

If the measurements pass the consistency test, we can combine
all three sets of data.  Even assuming no polarization information from
MAP, the result is that the errors on $w_g$ in the worst case of
$w_g=-1$ become $\sigma(w_g)=0.056$, allowing for a sharp test for the
presence of a cosmological constant.  Failure to achieve consistency
would indicate that one of our assumptions is wrong, e.g.
$w_g$ varies strongly with time or spatial curvature does not
vanish $\Omega_g \ne 1-\Omega_m$. 

Finally, note that by marginalizing $c_{\rm eff}$, our results
treat the clustering properties of the GDM as unknown and are thus
conservative in the context of scalar-field quintessence models 
\cite{WhiPrep}.  
For example if
$c_{\rm eff}$
is held fixed,
the limits on $w_g$ for the $w_g=-1/6$ model improve
by $\sim 30\%$ for MAP+SDSS with or without polarization; gains
are negligible near $w_g=-1$ since $c_{\rm eff}$ has little
observable effect there.

In summary, for any value of $w_g$, the combination of CMB data with SN
distance measures {\it or} galaxy surveys will provide reasonably
precise measures of $w_g$ and $\Omega_g$ even considering the unknown
clustering properties of the GDM.  The comparison of these two
combinations provides a sharp consistency test.  

\begin{table}[t]
\begin{tabular}{lcccccccc}
$c_{\rm eff}=0.03$ &
\multicolumn{2}{c}{MAP(P)+SDSS} &
\multicolumn{2}{c}{MAP(P)+SN} &
\multicolumn{2}{c}{Planck(P)+SDSS} &
\multicolumn{2}{c}{Planck(P)+SN} \cr
$w_g$ &
$\sigma(w_g)$ & $\sigma(\Omega_g)$ &
$\sigma(w_g)$ & $\sigma(\Omega_g)$ &
$\sigma(w_g)$ & $\sigma(\Omega_g)$ &
$\sigma(w_g)$ & $\sigma(\Omega_g)$ \cr
-1/6 &
0.015 (0.033) & 0.017 (0.028) &
0.024 & 0.030 &
0.009 (0.014) & 0.007 (0.010)&
0.014 & 0.009 \cr 
-1/3 &
0.027 (0.056) & 0.013 (0.022) &
0.028 & 0.031 &
0.016 (0.029) & 0.010 (0.019) &
0.020 & 0.013 \cr 
-1/2 &
0.047 (0.088) & 0.013 (0.022) &
0.041 & 0.034 &
0.022 (0.041) & 0.010 (0.019)&
0.021 & 0.011 \cr 
-2/3 &
0.074 (0.129) & 0.013 (0.022) &
0.063 & 0.037 &
0.029 (0.052) & 0.010 (0.020) &
0.023 & 0.010 \cr 
-5/6 &
0.108 (0.183) & 0.013 (0.022) &
0.091 & 0.040 &
0.037 (0.064) & 0.010 (0.019) &
0.026 & 0.009 \cr 
-1 &
0.126 (0.201) & 0.011 (0.018) &
0.125 & 0.042 &
0.033 (0.050) & 0.008 (0.016)&
0.027 & 0.010 \cr 
\end{tabular}
\caption{Errors on $w_g$ and $\Omega_g$ upon combining data sets.  
SDSS assumes information
out to $k_{\rm max}=0.2 h$Mpc$^{-1}$ or $0.1 h$Mpc$^{-1}$ (parentheses).
For $w_g < -1/6$, errors are insensitive to the sound speed of the 
fiducial model; here we use $c_{\rm eff}=\sqrt{0.03} \approx 0.2$.}
\label{tab:wgomg}
\end{table}

\begin{figure}
\centerline{ \epsfxsize=4.25truein \epsffile{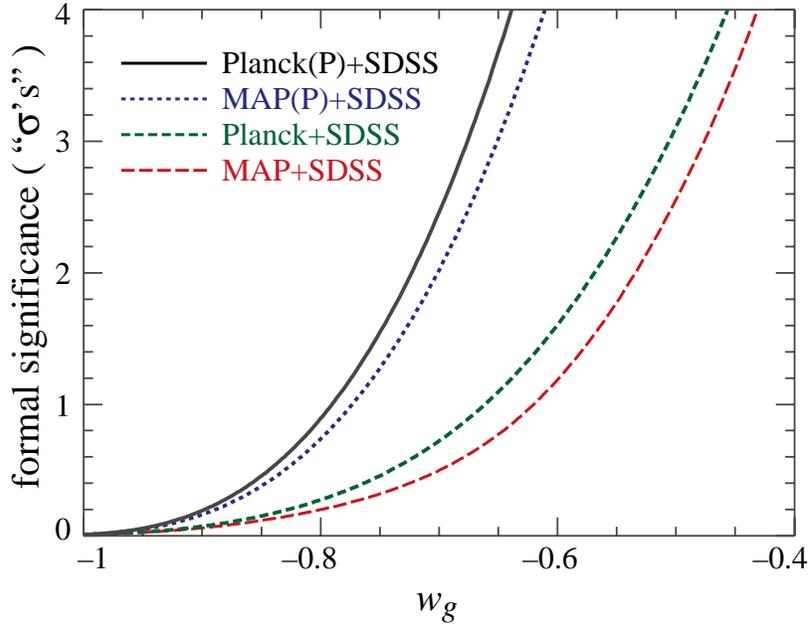}}
\caption{Significance of bounds on $c_{\rm eff}$ as a function of the
equation of state $w_g$ of the fiducial model.  Plotted is the formal
significance with which the simplest scalar-field models ($c_{\rm eff}=1$) may be excluded if the
true model is $c_{\rm eff}=\sqrt{0.03} \sim 0.2$. 
The formal errors may overestimate the significance as discussed in the text.
The power to distinguish sound speed effects
decreases sharply as one approaches the cosmological constant case of $w_g=-1$.
}
\label{fig:ceff}
\end{figure}
	
\section{Constraining the Sound Speed}
\label{sec:sound}

If the tests of the last section determine that the equation of state 
of the exotic component is in the range $-1 < w_g < 0$, we will have
discovered a new form of matter.   It then becomes interesting 
to explore its properties in order to search for a suitable
particle physics candidate. 
The simplest candidate is a slowly-rolling scalar field, also
known as ``quintessence'' \cite{Rat88,Sug92,Cob97,Cal98}. 
The hallmark of such a candidate 
is that its effective sound speed is simply the speed of light
($c_{\rm eff}=1$), i.e.\ it is a maximally stable form of matter.  
It furthermore has $c_{\rm vis}=0$.
Can the clustering properties of the GDM be measured well enough
to distinguish a scalar field component from alternate candidates
for the exotic matter? 

Stabilization of perturbations may occur through a finite
effective sound speed $c_{\rm eff}$ as it does for a real scalar
field or through viscosity $c_{\rm vis}$ as some defect-dominated 
models suggest \cite{Spe97,Buh98}.
Because the two are largely degenerate, 
we take $c_{\rm eff}$ as a proxy for both --
one actually determines the combination of $c_{\rm eff}$ 
and $c_{\rm vis}$ that fixes the GDM sound horizon.  

The sound speed
$c_{\rm eff}$ will only be well constrained if the GDM sound horizon at
$z_g$ is sufficiently small that cancellation of the ISW effect varies
strongly with $c_{\rm eff}$ (see Fig.\ \ref{fig:psound}) or if the
features in the matter power spectrum lie on scales accessible to
galaxy surveys.  Both of these considerations favor fiducial models 
with low $c_{\rm eff}$.  SN distance measures have no dependence on
$c_{\rm eff}$.  In Table \ref{tab:ceff} (upper), we show the errors
on $c_{\rm eff}$ as its fiducial value increases in an $w_g=-1/3$,
$\Omega_g=0.65$ fiducial model.  As usual, both these and other
parameters are marginalized when quoting errors on $c_{\rm eff}^2$.
As expected, the error $\sigma(c_{\rm eff}^2)$ increases sharply as
$c_{\rm eff}\rightarrow1$.

The strong variation of $\sigma(c_{\rm eff}^2)$ with $c_{\rm eff}^2$
itself makes it difficult to estimate the significance at which two
models can be separated.  For example, if we were to take
$\sigma(c_{\rm eff}^2=0.3)$ to infer that it is distinguishable from
$c_{\rm eff}^2=0$ at only $0.3/1.52=0.2\sigma$ from MAP+SDSS, we would
be incorrect since a model with a smaller value $c_{\rm eff}^2=0.03$ is
distinguishable from zero at a higher level.  Conversely, the ability
to distinguish a fiducial model with $c_{\rm eff}=0$ from one with
$c_{\rm eff} > 0$ is always overestimated.
This problem reflects the limitations of the Fisher matrix technique
caused by its infinitesimal expansion of the likelihood function.

To address this issue, we take an intermediate value of $c_{\rm eff}=\sqrt{0.03} 
\sim 0.2$
and ask how well we can reject the scalar
field hypothesis of $c_{\rm eff}=1$.    
We plot in Fig. \ref{fig:ceff}, the number of standard deviations 
by which the true sound speed is separated from the scalar field
value [``$\sigma$'s'' $=(1-c_{\rm eff}^2)/\sigma(c_{\rm eff}^2)]$.
Although this formal significance still overestimates the true
significance,  the qualitative result is clear.
If $w_g \simgt -1/2$, CMB and galaxy survey data will be able to place
interesting constraints on the sound speed. As $w_g$ decreases to $-1$,
the effects of clustering in the GDM vanish leaving no significant
constraint on $c_{\rm eff}$. A more complete exploration of the
likelihood function would yield more precise limits but is 
computationally time consuming; we defer such an analysis until
$w_g$ is measured and shown to be in this range.

\begin{table}
\begin{tabular}{ccccc} 
$c_{\rm eff}^2$ & \multicolumn{4}{c}{$\sigma(c_{\rm eff}^2)$} \cr
no priors: &  MAP+SDSS & MAP(P)+SDSS & Planck+SDSS & Planck(P)+SDSS \cr
0.03		&
0.13 (0.13)	&
0.06 (0.06)	&
0.12 (0.12)	&
0.04 (0.04)	\cr
0.1		&
0.46 (0.47)	&
0.14 (0.15)	&
0.42 (0.43)	&
0.10 (0.10) 	\cr
0.3 &
1.52 (1.55)	&
0.54 (0.54)	&
1.47 (1.47)	&
0.37 (0.37)  \cr
$\sigma(\ln\beta)=0.1$:&&&&\cr
0.03		&
0.06 (0.06)	&
0.06 (0.06)	&
0.06 (0.06)	&
0.04 (0.04)	\cr
0.1		&
0.18 (0.20)	&
0.14 (0.15)	&
0.18 (0.18)	&
0.10 (0.10) 	\cr
0.3 &
0.75 (0.83)	&
0.52 (0.53)	&
0.72 (0.72)	&
0.37 (0.38)  \cr
\end{tabular}
\caption{Errors on $c_{\rm eff}^2$ as a function of its fiducial value for
$w_g=-1/3$. 
SDSS assumes $k_{\rm max}=0.2 h$ Mpc$^{-1}$ ($0.1 h$ Mpc$^{-1}$).
P denotes the inclusion of CMB polarization information. Upper: no priors
on $\beta$; lower: prior of $\sigma(\ln\beta)=0.1$.   
The prior helps isolate $\tau$ if polarization information
is unavailable and hence brings the limits with and without polarization 
closer.} 
\label{tab:ceff}
\end{table}

\begin{figure}
\centerline{ \epsfxsize=3.25truein \epsffile{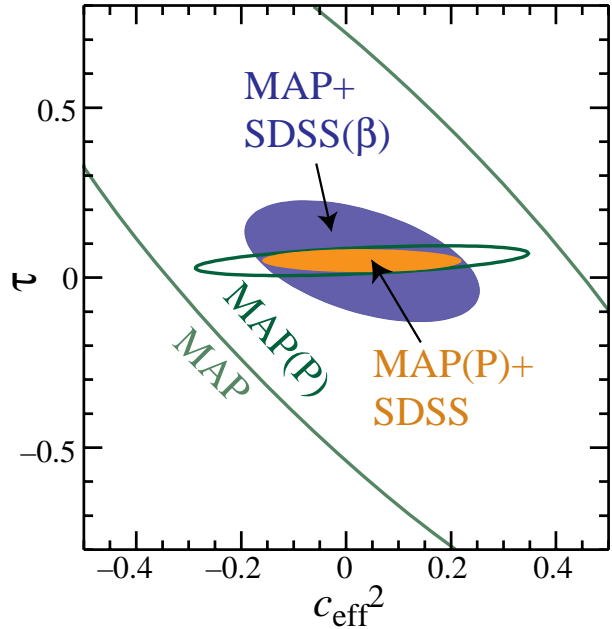}}
\caption{The $\tau-c_{\rm eff}^2$ degeneracy.  
$\tau$ and $c_{\rm eff}^2$ are nearly degenerate given only temperature
information.  With either polarization information on the CMB side [MAP(P)]
or a prior of $\sigma(\ln\beta)=0.1$ from redshift space distortions
on the galaxy power spectrum side [SDSS($\beta$)], 
the degeneracy is broken allowing better isolation of $c_{\rm eff}^2$.  
}
\label{fig:tauceff}
\end{figure}

Fig. \ref{fig:ceff} implies that good limits on $c_{\rm eff}$ depend
on polarization information.   This is because reionization
effects are nearly degenerate with ISW effects from $c_{\rm eff}$.
We show this in Fig.~\ref{fig:tauceff}.  Polarization information
isolates $\tau$ from the feature generated by Thomson scattering of
anisotropic radiation present at large scales during reionization.
However, because of foregrounds and systematics likely in the
large-angle polarization data, it is interesting to see whether any
other information can break this degeneracy.  The main effect of $\tau$
is to reduce the small-angle anisotropies in the CMB uniformly.  
If the intrinsic amplitude can be calibrated by the
galaxy survey data, $\tau$ could be measured.  With the growth function
and other transfer function effects under control, the remaining 
obstacle is the unknown bias factor $b$.  This can be measured on
large scales through redshift-space distortions.  
Since $\Omega_m$ is well-constrained by the
combination of CMB and galaxy survey data (\S\ \ref{sec:eos}), 
the constraint on $\beta=\Omega_m^{0.6}/b$ from these distortions 
supply information on the
bias.  Taking a conservative prior of $\sigma(\ln \beta)=0.1$
\cite{Hat98}, the normalization determination breaks the 
$c_{\rm eff}^2-\tau$ degeneracy almost as effectively as polarization
information.  We quantify this in Tab.~\ref{tab:ceff} (lower), where the
errors on $c_{\rm eff}^2$ with and without polarization are made more
comparable with this conservative prior on $\beta$.

In summary, interesting constraints on the clustering properties of the
exotic component will be available if $w_g \simgt -1/2$ as long as we
have either polarization data from the CMB or redshift-space distortion
information from galaxy surveys.

\section{Detecting the Neutrino Background Radiation}
\label{sec:nbr}

\begin{figure}
\centerline{ \epsfxsize=5.75truein \epsffile{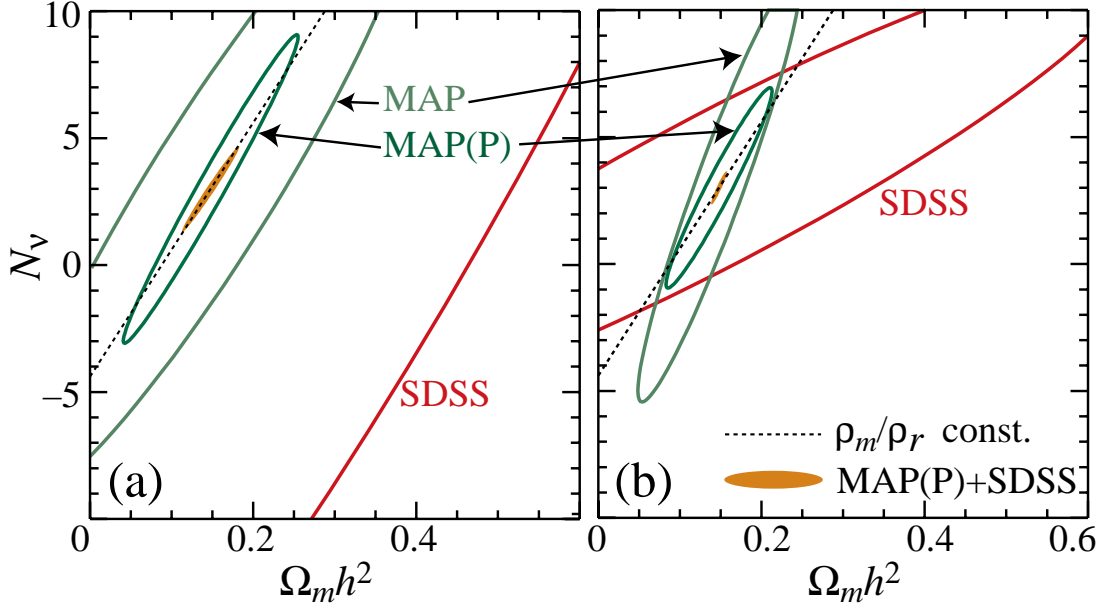}}
\caption{Matter-Radiation degeneracy.  
Plotted here are the 68\% CL for the various experiments.
(a) Ignoring the information supplied by the anisotropy of the NBR by
marginalizing over $c_{\rm vis}$, we find a strong degeneracy between
the neutrino number $N_\nu$ and the matter density $\Omega_m h^2$ along
a line of constant matter-radiation ratio $\rho_m/\rho_r$ in both the
CMB and galaxy survey data. Combining CMB information from MAP and
galaxy information from SDSS breaks the degeneracy somewhat. 
(b) Utilizing the information in the anisotropies by assuming 
$c_{\rm vis}=\sqrt{1/3}$ shrinks and rotates the error ellipses leading to better
complementarity and tighter errors.  }
\label{fig:nnu}
\end{figure}

If the equation of state of the GDM is determined to be $w_g=-1$, then 
the only possibility is a cosmological constant.  In this case,
the basic aspects of structure formation are so simple that subtle effects
in the dark sector can be uncovered.  The remaining dark matter
in the universe is the neutrino background radiation (NBR).  How 
well can we detect its presence?  The issue is somewhat more subtle than it
initially appears due to a degeneracy in the CMB acoustic peaks.  We shall
see that detecting the neutrino background radiation requires detecting its
fluctuations, in particular its anisotropies.

\subsection{Matter-Radiation Degeneracy}

Given that CMB anisotropies are generally sensitive to
changes in the expansion rate at high redshift,  
one might think the radiation content of the universe could be measured 
precisely.
Indeed the matter density $\Omega_m h^2$ can be measured
to $\sigma(\Omega_m h^2) = 0.02$ by the MAP satellite (without polarization) if
the radiation is taken to be fixed.
The problem is that what the CMB best measures
is the matter-radiation ratio, not the matter or radiation density 
individually.   

The GDM parameterization can be used to explore this
degeneracy and more generally deconstruct 
the information contained about the NBR in the CMB.  
We know that the matter-radiation degeneracy arises because
of the way the background expansion rate scales with these parameters
[see Eq.~(\ref{eqn:hubble})].   Fluctuations in the matter and
radiation break this degeneracy.  We can use the GDM parameterization
to separate the information on the background expansion rate
from that of the fluctuation properties.  

As shown in \S \ref{sec:gdm} and \cite{Hu98}, 
the NBR is accurately modeled by a GDM
component with $w_g=c_{\rm eff}^2=c_{\rm vis}^2=1/3$ and density
$\Omega_g = 5.63 \times 10^{-6} h^{-2} N_\nu$.  $N_\nu=3$ in the
fiducial model.  Thus $N_\nu$ and $w_g$ determine the background
properties whereas $c_{\rm eff}$ and $c_{\rm vis}$ control the
fluctuation properties, with $c_{\rm vis}$ controlling the
anisotropic stress of the NBR.  The anisotropic stress is proportional
to the quadrupole anisotropy of the NBR.  Thus $c_{\rm vis}=0$ (with
$c_{\rm eff}=\sqrt{1/3}$) represents a component with the same background
properties as the NBR but with no anisotropies.

How does ignorance of the properties of the NBR affect the determination of
the matter density $\Omega_m h^2$?  
If we allow $N_\nu$ and $c_{\rm vis}$ to vary so as to eliminate the information
provided by the density and anisotropy of the radiation component, the
error ellipses of Fig.~\ref{fig:nnu}a
reveal a matter-radiation degeneracy, 
i.e.\ they are elongated along the line of constant matter-radiation density
ratio.
The degeneracy affects both CMB measurements and galaxy surveys alike.
Here and in the remainder of this section, we keep 
$w_g$ and $c_{\rm eff}$ fixed while varying $\Omega_g$, $c_{\rm vis}$, and
the other cosmological parameters including a cosmological constant
$\Omega_\Lambda$.
The MAP errors on $\Omega_m h^2$ are degraded from 
$\sigma(\Omega_m h^2)=0.02$ to $0.16$ when $N_\nu$ is allowed to vary.
The baryon-to-photon ratio $\Omega_b h^2$ remains well-measured.
Some leverage in a flat universe is provided by the fact that the
actual matter density comes into the angular diameter distance for the CMB. 
Indeed, if one fixes the other parameters that go into the angular
diameter distance, in this context $\Omega_\Lambda$, then the CMB does
place tight constraints on $\Omega_m h^2$ and $N_\nu$  separately \cite{Lop98}.  However,
in the
general case, the angular diameter distance degeneracy prevents a
precise measurement of $\Omega_m h^2$ by these means.

Combining CMB anisotropies and galaxy power spectrum information, which
both suffer from the matter-radiation degeneracy individually, restores
tight error bars on $\Omega_m h^2$ even using only temperature
information from MAP (see Tab. \ref{tab:Nnu}).  Here the additional
information on $\Omega_b/\Omega_m$ from baryonic features in the galaxy
power spectrum along with the precise measurement of
$\Omega_b h^2$ from the CMB constrains $\Omega_m h^2$.
This is another example in which the complementary nature of the CMB and
galaxy survey data helps in a subtle way.  

\subsection{Limiting the Number of Neutrinos}

In the standard scenario, the fluctuations of the NBR are not unknown;
they are fixed through the properties of the neutrinos and gravitational instability. 
These fluctuations in the NBR further breaks the
matter-radiation degeneracy.  Like the CMB itself, the NBR carries
temperature anisotropies (see \cite{Hu95} for the full angular power
spectrum).  In particular, the quadrupole anisotropy of neutrinos
alters the gravitational potentials that drive acoustic oscillations
(see Fig.\ \ref{fig:pcvis}).

By fixing $c_{\rm vis}=\sqrt{1/3}$, we can ask how well the number $N_\nu$
of neutrino species in the NBR can be measured under the standard
assumptions. 
This is appropriate for either flavored or sterile 
neutrinos if their mass is sufficiently small ($m_\nu\lesssim0.1$ eV).
We show the results in Fig.~\ref{fig:nnu}b and Tab.~\ref{tab:Nnu}.  The
CMB and galaxy survey ellipses in the $N_\nu$--$\Omega_m h^2$ plane
shrink and rotate in opposite senses from the case of a marginalized
$c_{\rm vis}$ (Fig.~\ref{fig:nnu}a).  This enhances the complementary
nature of CMB and galaxy survey data giving a substantial improvement
when the data sets are combined if only MAP data is available.

As the errors in Tab.\ \ref{tab:Nnu} are comparable to those achievable
from big bang nucleosynthesis (BBN) \cite{Sch98}, CMB and galaxy
surveys should provide a powerful consistency check on BBN and
a constraint on additional neutrino species populated around
recombination.  Unfortunately, testing
percent-level differences in the NBR temperature (represented here by a
change $\delta N_\nu/N_\nu = 4\delta T_\nu/T_\nu$, see
Eq.~[\ref{eqn:omeganh2}]) due to the details of their decoupling
\cite{Dod92,Lop98,Gne98} seems out of reach even if we combine all of our
precision tests.

\begin{table}
\begin{tabular}{lcccccccc} 
   	       & \multicolumn{2}{c}{MAP+SDSS}
	       & \multicolumn{2}{c}{MAP(P)+SDSS}
	       & \multicolumn{2}{c}{Planck+SDSS}
	       & \multicolumn{2}{c}{Planck(P)+SDSS} \cr
assumption     & $\sigma(\Omega_m h^2)$ & $\sigma(N_\nu)$
	       & $\sigma(\Omega_m h^2)$ & $\sigma(N_\nu)$
               & $\sigma(\Omega_m h^2)$ & $\sigma(N_\nu)$
               & $\sigma(\Omega_m h^2)$ & $\sigma(N_\nu)$ \cr
unknown $c_{\rm vis}^2$		&
0.026 (0.046) & 1.24 (2.26) &
0.023 (0.036) & 1.12 (1.92) &
0.006 (0.006) & 0.30 (0.43) &	
0.004 (0.004) & 0.21 (0.23) \cr
fixed $c_{\rm vis}^2$		&
0.007 (0.024) & 0.44 (1.59) &
0.006 (0.022) & 0.43 (1.44)  &	
0.003 (0.005) & 0.23 (0.43) &
0.003 (0.003) & 0.17 (0.20) \cr
\end{tabular}
\caption{Errors on $\Omega_m h^2$ and $N_\nu$ with and without information from NBR
anisotropies.
SDSS assumes $k_{\rm max}=0.2 h$ Mpc$^{-1}$ ($k_{\rm max}=0.1 h$ Mpc$^{-1}$).
P denotes the inclusion of CMB polarization information.}
\label{tab:Nnu}
\end{table}

\subsection{Detecting Neutrino Anisotropies}

Anisotropies in the NBR are predicted by the gravitational instability
paradigm and are potentially observable through their effect on CMB
anisotropies.  In the GDM model for the NBR, the anisotropies are
determined by the viscosity parameter $c_{\rm vis}$; constraints
on this parameter tell us how well anisotropies in the NBR can be
detected.  
The role of the viscosity parameter $c_{\rm vis}$ in breaking
degeneracies in the last section suggests that the
anisotropies in the neutrino background radiation may themselves be
detectable.  
Unfortunately, with the CMB alone, the effect is strongly degenerate
with those of other parameters. Even though we fix the other properties
of the GDM ($w_g=c_{\rm eff}^2=1/3$, $N_\nu=3$), changes in 
$c_{\rm vis}$ can be mimicked by changes in 
the normalization and tilt of the spectrum at
small angles.  

\begin{figure}
\centerline{ \epsfxsize=3.25truein \epsffile{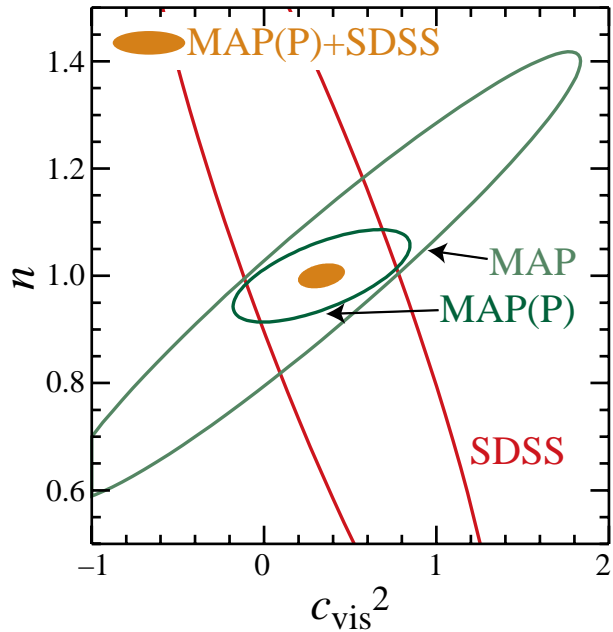}}
\caption{Detecting anisotropies in the neutrino background radiation.
Anisotropies in the NBR are detectable if a model with 
$c_{\rm vis}^2=1/3$ can be distinguished from one with $c_{\rm vis}^2=0$.
Degeneracy with other parameters such as the tilt $n$ prevent the CMB
information provided by MAP from detecting the anisotropies.  Adding
galaxy survey information from SDSS breaks the degeneracy and allows a
statistically significant detection.  We have assumed here that $N_\nu$
is fixed at 3.}
\label{fig:cvis}
\end{figure}

By adding in galaxy survey data, $c_{\rm vis}^2=1/3$
(NBR anisotropies) and $c_{\rm vis}^2=0$ (no anisotropies) are
separated by $3.5\sigma$ from MAP+SDSS.  The significance improves to
8.7$\sigma$ with Planck.

How sensitive is the measurement to the underlying assumptions about
the data set and model space?  The loss of polarization information
does not significantly affect these limits.  On the other hand, if we
take the more conservative $k_{\rm max}=0.1h$Mpc$^{-1}$ for the galaxy
surveys, the significance decreases to $1.2\sigma$ ($7.3\sigma$) for
MAP+SDSS (Planck+SDSS).  Perhaps more important, the MAP+SDSS result
does depend on prior knowledge that $N_\nu \approx 3$.  Fortunately,
even assuming only very conservative constraints of $\sigma(N_\nu)=1.0$
from big bang nucleosynthesis, allows a detection at 2.0$\sigma$
(7.1$\sigma$) for MAP+SDSS (Planck+SDSS).  These results imply that NBR
anisotropies can be detected with high significance at least by the
Planck satellite, even under conservative assumptions.

\section{Conclusions}
\label{sec:conclusions}

With the wealth of precision cosmological measures that are becoming
available, we should soon be in the position to identify all of the
cosmologically important components of the universe -- including any
dark components that may be present. Not only do CMB anisotropies,
high-redshift objects, and galaxy surveys probe different aspects of
the cosmology, but they can work together to uncover a
complete and consistent cosmological picture.

We have shown here how the combined power of these data sets can
determine the properties of the dark components.   Assuming that the
preliminary indications from SN data that the missing component is not
merely spatial curvature are confirmed, the first step will be to
determine the equation of state of the exotic component.  The task is
non-trivial due to a degeneracy with its density in determining the
expansion rate.  The degeneracy is broken by combining the CMB data
with SN distance measures, galaxy surveys, or any other measurement
that can constrain $\Omega_m=1-\Omega_g$ or $h$.  By further combining
any of these pairs of data sets, we create powerful consistency
tests.   Note that these tests work even near a cosmological constant model
$p_g/\rho_g \equiv w_g =-1$, where the degeneracy in the CMB alone is at
its worst.

Should the equation-of-state measurement rule out a cosmological
constant, we will want to study the other properties that identify this
exotic component.  Its clustering scale is accessible in the CMB and
galaxy surveys as features in their power spectra at large scales.  We
have parameterized this with a sound speed and have shown that as long
as the equation of state is sufficiently different from a cosmological
constant ($w_g \simgt -1/2$), one can distinguish a maximally stable
component ($c_{\rm eff}=1$) from a component with $c_{\rm eff} \simlt
0.2$.  Distinguishability increases substantially as $p_g/\rho_g$ increases,
reaching $c_{\rm eff} \simlt 0.6$ at $w_g=-1/6$.  Such limits are
interesting since the simplest physically motivated exotic component, a
slowly-rolling scalar-field quintessence, is maximally stable
with $c_{\rm eff}=1$.

We have only considered models where the equation of state 
varies in time sufficiently
slowly to be replaced by some suitably averaged but constant $w_g$.
If the GDM sector involves stronger temporal variation, 
to what extent will upcoming data sets be able to constrain the
possibilities?  As noted in \S \ref{sec:eos},
acoustic features in the CMB anisotropies when combined with
those in the galaxy power spectrum can be combined to measure
$\Omega_m$.  This measurement requires that the dark components at 
high redshift such as the NBR are known but makes no assumptions
whatsoever about the low-redshift behavior of $w_g$.  
With the present-day value of $\Omega_g=1-\Omega_m$ 
known to fair accuracy, there
are a number of observational handles on $w_g$ as a function of time.
The location of the CMB acoustic peaks determines the angular diameter
distance to high redshift.  Mid-redshift supernovae constrain the
luminosity distance to $z\approx0.5$, although with very large samples
one may even extract some redshift dependence. 
Measurements of the normalization of the power spectrum on scales
below the GDM sound horizon will
constrain the GDM-modified growth rate.  The $z=0$ normalization may be
measured from abundances of rich clusters \cite{Lim98}
and from the galaxy power
spectrum given a measurement of galaxy bias from redshift distortions
or peculiar velocity data sets.  The normalization at higher redshift
can be estimated from the statistics of the Lyman $\alpha$ forest,
damped Lyman $\alpha$ systems, and high-redshift clusters.  Hence,
although the most general equation of state is described by a free function
of redshift, 
there are actually a number of robust observational handles on its behavior!

A similar analysis shows that even if the universe contains both
an accelerating component and a non-vanishing spatial curvature, which in many
respects resembles a universe with $w_g$ varying from $-1/3$ to $-1$, 
the combination of information from different redshifts will give us 
leverage on the two separately.  The situation is actually even more
favorable since the geometrical aspects of curvature enter strongly into
the angular diameter distance as measured by the CMB.

Should the measurement of the equation of state confirm the 
relative simplicity of a cosmological
constant, we will be able to probe in detail the remaining dark
component, the neutrino background radiation.  Detection of the
neutrino background radiation through the CMB suffers from the fact
that a change its energy density may be compensated with a change in
the matter density up to effects due to the presence of fluctuations.
We have shown that by combining CMB and galaxy survey data, this
degeneracy can be broken and yields constraints competitive with those
from big bang nucleosynthesis.  Furthermore, the combination of CMB and
galaxy survey data should provide the first detection of anisotropies
in the neutrino background radiation.  The detection of these
anisotropies, predicted by the gravitational instability paradigm for
structure formation, would represent a triumph for cosmology.

\smallskip
\noindent{Acknowledgements:}  
W.H.\ is supported by the Keck Foundation and a Sloan Fellowship,
W.H.\ and D.J.E. by NSF-9513835,
D.J.E.\ by a Frank and Peggy Taplin Membership at the IAS, and 
M.T.\ by NASA through grant NAG5-6034 and Hubble
Fellowship HF-01084.01-96A from STScI, operated by AURA, Inc.
under NASA contract NAS4-26555.


\begin{thebibliography}{99}

\bibitem{Map}
        MAP: http://map.gsfc.nasa.gov

\bibitem{Planck}
  	Planck: http://astro.estec.esa.nl/SA-general/Projects/Planck

\bibitem{Jun96}
        G. Jungman, M. Kamionkowski, A. Kosowsky, \& D. N. Spergel, Phys. Rev. D
        {\bf 54}, 1332 (1996) 
	[astro-ph/9512139];  
        J.R. Bond, G. Efstathiou, \& M. Tegmark, Mon. Not. Roy. Astr. Soc. {\bf 291}, L33 (1997) 
	[astro-ph/9702100]; 
	M. Zaldarriaga, D. N. Spergel, \& U. Seljak, Astrophys. J. {\bf 488}, 1 
 (1997)
	[astro-ph/9702157]

\bibitem{Per97}
        S. Perlmutter, et al., Astrophys. J. {\bf 483}, 565 (1997) 
	[astro-ph/9608192];
	S. Perlmutter, et al. Nature (London) {\bf 391}, 51 (1998)

\bibitem{Gar97}
	P. M. Garnavich, et al., Astrophys. J. Lett {\bf 493}, 53 (1998) 
	[astro-ph/9710123];
	A. G. Riess, et al., preprint [astro-ph/9805201]

\bibitem{2dF}
	2dF: http://meteor.anu.edu.au/$\sim$colless/2dF

\bibitem{SDSS}
	SDSS: http://www.astro.princeton.edu/BBOOK


\bibitem{Hu98}
	W. Hu, Astrophys. J. (to be published) [astro-ph/9801234]

\bibitem{Hue98}
        G. Huey, L. Wang, R. Dave, R.R. Caldwell, \& P. J. Steinhardt, 
	preprint [astro-ph/9804285]

\bibitem{Rat88}
	B. Ratra \& P. J. E. Peebles, Phys. Rev. D. {\bf 37}, 3406 (1988);
	P. J. E. Peebles \& B. Ratra, Astrophys. J. {\bf 325}, L17 (1988)

\bibitem{Sug92}
	N. Sugiyama \& K. Sato, Phys. Rev. D. {\bf 387}, 439 (1992) 

\bibitem{Cob97}
	J. A. Frieman, C. T. Hill, A. Stebbins, \& I. Waga, 
	Phys. Rev. Lett. {\bf 75}, 2077 (1995) 
	[astro-ph/9505060];
        K. Coble, S. Dodelson, \& J. Frieman, Phys. Rev. D {\bf 55}, 1851 (1997) 
	[astro-ph/9608122];
	J. A. Frieman \& I. Waga, Phys. Rev. D
	{\bf 57}, 4642 (1998)
	[astro-ph/9709063]

\bibitem{Cal98}
        R. R. Caldwell, R. Dave, \& P. J. Steinhardt, Phys. Rev. Lett. {\bf 80} 1582 (1998)
        [astro-ph/9708069]; 
	R. R. Caldwell \& P. J. Steinhardt, Phys. Rev. D. {\bf 57}, 6057 (1998) 
	[astro-ph/9710062]


\bibitem{Kod84}
        H. Kodama \& M. Sasaki, M. Prog. Theor. Phys. Supp. {\bf 78}, 1 (1984)


\bibitem{Chi97}
        T. Chiba, N. Sugiyama, \& T. Nakamura, Mon. Not. Roy. Astr. Soc. {\bf 289} 5 (1997) 
        [astro-ph/9704199];
        T. Chiba, N. Sugiyama, \& T. Nakamura, preprint
	[astro-ph/9806332]
	

\bibitem{Tur97}
        M. S. Turner \& M. White, Phys. Rev. D. {\bf 56}, 4439 (1997)
        [astro-ph/9701138]



\bibitem{Spe97}
	D. N. Spergel \& U.-L. Pen, Astrophys. J. Lett. {\bf 491}, L67 (1997)
	[astro-ph/9611198]

\bibitem{Buh98}
	M. Bucher, private communication

\bibitem{defects} 
	N. Turok, U.-L. Pen, \& U. Seljak, preprint [astro-ph/9706250]

\bibitem{CalTol}
	M. Hamuy, et al., Astronomical J. {\bf 112}, 2398 (1996)
	[astro-ph/9609064]

\bibitem{Whi98}
	M. White, Astrophys. J. (to be published)
	[astro-ph/9802295]

\bibitem{Fix96}
	D. Fixsen, et al. Astrophys. J. {\bf 473}, 576 (1996)
	[astro-ph/9605054] 

\bibitem{Eis98}
	D. J. Eisenstein, W. Hu, \& M. Tegmark, Astrophys. J. (submitted) 

\bibitem{Teg97a}
	M. Tegmark, A. N. Taylor, \& A. F. Heavens, Astrophys. J. {\bf 480}, 22 (1997)
	[astro-ph/9603021]

\bibitem{Sel96}
	U. Seljak \& M. Zaldarriaga, Astrophys. J. {\bf 469} 437 (1996)
	[astro-ph/9603033]

\bibitem{Hu95}
	W. Hu, D. Scott, N. Sugiyama, \& M. White, Phys. Rev. D.  {\bf 52}, 5498 (1995)
	[astro-ph/9505043]

\bibitem{Whi96}
	M. White \& D. Scott, Astrophys. J. {\bf 459}, 415 (1996) 
	[astro-ph/9508157]

\bibitem{Hu97d}
	W. Hu \& M. White,  Astrophys. J. {\bf 479}, 568 (1997)
	[astro-ph/9609079] 	

\bibitem{Pre92}
	W. H. Press, S. A. Teukolsky, W. T. Vetterling, \& B. P. Flannery, 
	{\it Numerical Recipes in Fortran, Second Edition}
	(Cambridge University Press, New York, 1992)

\bibitem{Teg97b}
	M. Tegmark, Phys. Rev. Lett. {\bf 79}, 3806 (1997)
	[astro-ph/9706198]	

\bibitem{Mei98}
	A. Meiksen, M. White, \& J. Peacock (in preparation)

\bibitem{Teg98}
	M. Tegmark, D. J. Eisenstein, W. Hu, \& R. Kron, preprint 
	[astro-ph/9805117]

\bibitem{WhiPrep}
	M. White (in preparation)

\bibitem{Eis98L}
	D. J. Eisenstein, W. Hu, \& M. Tegmark, preprint 
	[astro-ph/9805239]

\bibitem{Hu96}
	W. Hu \& M. White, Astrophys. J. 471, 30 (1996)
	[astro-ph/9602019]

\bibitem{Hat98}
	S. Hatton \& S. Cole, Mon. Not. Roy. Astron. Soc. {\bf 296}, 10 (1998)
	[astro-ph/9707186]

\bibitem{Dod92}
	S. Dodelson \& M. Turner, Phys. Rev. D. {\bf 46}, 3372 (1992)

\bibitem{Lop98}
        R. E. Lopez, S. Dodelson, A. Heckler, \& M.S. Turner, preprint
	[astro-ph/9803095]

\bibitem{Gne98}
	N. Y. Gnedin \& O. Y. Gnedin, preprint 
	[astro-ph/9712199]

\bibitem{Sch98}
	D. N. Schramm \& M. S. Turner, Rev. Mod. Phys. 70, {\bf 303} (1998) 
	[astro-ph/9706069]

\bibitem{Lim98}
	L. Wang \& P. J. Steinhardt, preprint 
	[astro-ph/9804015] 

\end{thebibliography}
\end{document}